\documentclass[prb,twocolumn,showpacs,amsmath,amssymb]{revtex4-1}
\usepackage{amssymb}
\usepackage{graphicx}
\usepackage{bm}
\usepackage{amsmath}
\renewcommand{\v}[1]{{\bf #1}}

\newcommand{\beq}{\begin{equation}}
\newcommand{\eeq}{\end{equation}}
\newcommand{\beqn}{\begin{eqnarray}}
\newcommand{\eeqn}{\end{eqnarray}}

\renewcommand{\vec}[1]{\mbox{\boldmath$#1$}}
\begin{document}

\title{Universal spin-triplet superconducting correlations of Majorana fermions}
\author{Xin Liu} \email{phyliuxin@gmail.com}
\author{Jay D. Sau} 
\author{S. Das Sarma}
\affiliation{Condensed Matter Theory Center and Joint Quantum Institute, Department of Physics, University of Maryland, College Park, Maryland 20742-4111, USA}
\date{\today}

\begin{abstract}
We establish that Majorana fermions on the boundary of topological superconductors have only spin-triplet superconducting correlations independent of whether the bulk superconducting gap is spin singlet or triplet. This is universal for time-reversal broken (respected) topological superconductors (TSCs) with an odd number of (pairs of) Majorana fermions on the boundary. Consequently, resonant Andreev reflection induced by Majorana fermions only occurs in spin-triplet channels and always injects spin-triplet Cooper pairs into the leads. This spin-triplet condensate results in a spin-orbit coupling (SOC) controlled oscillatory critical current without a $0$-$\pi$ transition in the TSC/SOC-semiconductor/TSC Josephson junction. The observation of this unique current-phase relation can serve as the definitive signal for Majorana fermions. Our study shows a technique for manipulating Majorana fermions based on their spin-triplet superconducting correlations.

\end{abstract}

\maketitle

{\it Introduction--} Due to recent theoretical proposals of realizing topological superconductors (TSCs) with an $s$-wave superconducting proximity effect in spin-orbit coupling (SOC) systems \cite{TSC:Fu2008,TSC:Zhang2008,TSC:Sato2009,TSC:Sau2010,TSC:Lutchyn2010,TSC:Oreg2010,TSC:Sau2010a,TSC:Alicea2010,TSC:Zhang2013,TSC:Ebisu2014,TSC:Liu2014,TSC:Hui2015,
TSC:Brydon2015}, finding and manipulating Majorana fermions \cite{TSC:Read2000,TSC:Kitaev2001,TSC:Nayak2008} have become possible in superconductor-semiconductor devices \cite{TSC:Mourik2012,TSC:Deng2012,TSC:Rokhinson2012,TSC:Das2012,TSC:Wang2012,TSC:Churchill2013,TSC:Xu2014}. Majorana fermions (MFs) in this context are symmetry protected zero-energy subgap states of TSCs with equal weight electron-hole superposition. This leads to some unique transport properties, such as quantized zero-bias conductance \cite{TSC:Sengupta2001,TSC:Law2009,TSC:Flensberg2010} and fractional Josephson effect \cite{TSC:Lutchyn2010,TSC:Kitaev2001,TSC:Kwon2004,TSC:Fu2009,TSC:Zhang2014,TSC:Ebisu2014}, and exotic superconducting condensates such as an odd frequency $s$-wave and spin-triplet pairing in a time-reversal broken $p_x$-wave-superconductor/normal-metal junction \cite{TSC:Asano2013}. However, so far, most studies ignore the spin states of MFs. On the other hand, controlling or manipulating spin states is a conceptional revolution in fundamental science as well as emerging technologies \cite{Zutic:2004_a} and is triggering important fundamental discoveries of spin-dependent transport phenomena such as giant magnetoresistance \cite{Baibich:1988,Binash:1989} and spin Hall effect \cite{Murakami:2003_a,Sinova:2004_a}. Therefore, it is natural to ask whether we can manipulate MFs by their spin properties.

In this paper, we show that MFs, zero-energy bound states of TSCs, have only spin-triplet superconducting correlations which are model independent and guaranteed by particle-hole symmetry for time-reversal broken TSCs (D class \cite{TSC:Schnyder2008}) and particle-hole as well as time-reversal symmetries for time-reversal respected TSCs (DIII class \cite{TSC:Schnyder2008}). This leads to our experimental prediction that no matter whether the bulk superconducting gap is spin singlet \cite{TSC:Law2009,TSC:Sau2010a} or triplet \cite{TSC:Tanaka2005,TSC:Tanaka2007,TSC:Flensberg2010}, the MF-induced resonant Andreev reflection (AR) only injects spin-triplet Cooper pairs into the normal lead. We perform a symmetry analysis as well as numerical calculations of the reflection matrix at the interface of either time-reversal broken or respected TSC/normal-metal (NM) junctions. We further propose to detect spin-triplet Cooper pairs in a TSC/SOC-semiconductor/TSC experimental setup by observing a SOC controlled oscillatory critical Josephson current. It has been recently shown that SOC can rotate the $\bm{d}$ vector of the spin-triplet Cooper pairs and leave spin-singlet Cooper pairs unaffected in the proximity region. Therefore, the observation of our predicted SOC controlled critical current can serve as direct confirmation for spin-triplet Cooper pairs and the MF transport signal.

{\it Spin-triplet superconducting correlations of MFs} In any dimension superconductor, the MF, a self-Hermitian quasiparticle, can be defined as $\gamma_{i}(\vec{x})=\left(\psi_{\uparrow},\psi_{\downarrow},\psi^{\dagger}_{\downarrow},-\psi^{\dagger}_{\uparrow}\right)\Psi_{i}(\vec{x})$, where $\vec{x}$ is generally a vector and $\Psi_{i}(\vec{x})$ is the $i$th zero-energy quasiparticle wave function. Electron and hole correlations at the Fermi surface show up in the block off-diagonal part of the equal space zero energy spectral function, which is a Hermitian matrix and has the form $A(E=0,\vec{x})=\sum_i |\Psi_i(\vec{x})\rangle \langle \Psi_i(\vec{x})|$ \cite{Suppl}. In the four-component spinor basis $\left(\psi_{\uparrow},\psi_{\downarrow},\psi^{\dagger}_{\downarrow},-\psi^{\dagger}_{\uparrow}\right)^{\rm T}$, the block off-diagonal part of the spectral function (BOSF) can be written as
\begin{eqnarray*}
A^{\rm off}(E,\bm x)=\left(\begin{array}{cc} \hat{0} & (d_{0}\sigma_0+{\vec d}\cdot \bm{\sigma})  \\ \left[(d_{0}\sigma_0+{\vec d}\cdot \bm{\sigma}) \right]^{\dagger} & \hat{0} \end{array}\right),
\end{eqnarray*} 
where $\sigma_{0}$ and $\vec{\sigma}$ are the $2 \times 2$ identity and Pauli matrices in electron-hole and spin space respectively, and $d_0$ and \vec{d} are the spin-singlet and spin-triplet pairing amplitudes, respectively.
In a superconductor, the particle-hole symmetry requires the zero energy spectral function to satisfy \cite{Suppl} \begin{eqnarray}\label{A-ph}
\hat{C} A(E=0) \hat{C}^{-1}=A(E=0),
\end{eqnarray}
where $\hat{C}=\tau_y\sigma_y K$ is the particle-hole operator, with $K$ the complex conjugate operator and $\tau_{x,y,z}$ the Pauli matrices in particle-hole space. A straightforward calculation shows that 
\begin{eqnarray}\label{pair-ph}
&&\hat{C}A^{\rm off}\hat{C}^{-1}=\tau_y\sigma_y (A^{\rm off})^* \tau_y\sigma_y \nonumber \\ &&=\left(\begin{array}{cc} \hat{0} & (-d_0\sigma_0+{\vec d}\cdot \rm{\sigma})  \\ \left[(-d_0\sigma_0+{\vec d}\cdot \rm{\sigma}) \right]^{\dagger} \end{array}\right).
\end{eqnarray} 
Comparing Eq.~(\ref{pair-ph}) with Eq.~(\ref{A-ph}), we find that $d_0$ must vanish at the Fermi surface. This implies that the particle-hole symmetry completely forbids spin-singlet pairing but allows spin-triplet pairing in the zero-energy BOSF. For the D class TSC, when there is one MF $\gamma_1=c_{\kappa}+c_{\kappa'}^{\dagger}$ on the boundary, the Majorana condition $\gamma_1^{\dagger}=\gamma_1$ requires $\kappa=\kappa'$, which results in a spin-triplet BOSF, $A^{\rm off}=(\tau_x\sigma_x-\tau_y\sigma_y)/2$, by choosing $\kappa= \uparrow$. 
If there is an additional zero mode, $\gamma_2=i(c_{\kappa}-c_{\kappa}^{\dagger})$, the anomalous spectral function is zero because the two MFs can emerge as a single electron or hole \cite{Suppl}. For a DIII class TSC, although there is always at least a pair of MFs on the boundary, the time-reversal symmetry guarantees that these two MFs belong to different spin channels and therefore manifest spin-triplet pairing \cite{Suppl}. The above analysis applies equally well to the multi-MF situation and we conclude that an odd number (of pairs) of MFs in the D (DIII) class have only spin-triplet superconducting correlations. It is noted that our symmetry analysis is independent of the details of TSC models and robust against the spin-independent disorder. Therefore the spin-triplet superconducting correlation is a universal and robust property of MFs.  

\begin{figure}[ht]
\centering
\begin{tabular}{l}
\includegraphics[width=0.8\columnwidth]{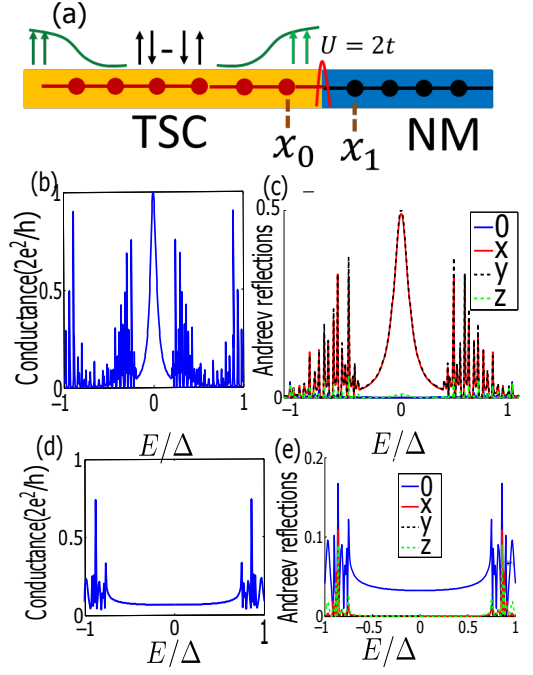}
\end{tabular}
\caption{(Color online) (a) Time-reversal broken TSC/NM junction. The black arrows represent the spin-singlet bulk superconducting gap. The green arrows represent the MF superconducting correlations. The green curves indicate the spatial distribution of MFs. The red peak indicates the location of the barrier with the potential $ U = 2 t$. (b) (c) The superconductor is in the topological nontrivial regime. The conductance and AR probabilities in spin-singlet and spin-triplet channels as a function of the energy inside the superconducting gap.  (d), (e) The superconductor is in the topological trivial regime. The conductance and AR probabilities in spin-singlet and spin-triplet channels as a function of the energy inside the superconducting gap. In (c) and (e), the solid blue, solid red, dashed black and dashed green lines represent AR probabilities, $|a_0|^2$, $|a_x|^2$, $|a_y|^2$, and $|a_z|^2$ for spin-singlet and spin-triplet channels respectively.}
\label{TSC-D}
\end{figure}

{\it MF-induced resonant spin-triplet Andreev reflections} 
The current experimental methods for detecting MFs \cite{TSC:Mourik2012,TSC:Deng2012,TSC:Rokhinson2012,TSC:Das2012, TSC:Churchill2013} are related to ARs. Therefore, to experimentally confirm our prediction of the universal spin-triplet superconducting correlations of MFs, we first focus on the AR in spin-singlet and spin-triplet channels which we discuss below. The scattering matrix at the interface of a one-dimensional (1D) SC/NM junction generally has the form \cite{TSC:Beenakker2011,TSC:Beenakker2014} \begin{eqnarray}\label{AR-1}
\hat{R}(E)&=&\left(\begin{array}{cc} \hat{r}_{\rm ee}(E)  & \hat{r}_{\rm eh}(E) \\ \hat{r}_{\rm he}(E) & \hat{r}_{\rm hh}(E) \end{array}\right),\nonumber \\
\hat{r}_{\rm eh}(E)&=&a_{0}\sigma_{0}+a_{j}\sigma_{j}=-\sigma_y\hat{r}^*_{\rm he}(-E)\sigma_y,\nonumber \\
\hat{r}_{\rm ee}(E)&=&b_{0}\sigma_{0}+b_{j}\sigma_{j}=\sigma_y\hat{r}^*_{\rm hh}(-E)\sigma_y,
\end{eqnarray}
where $\hat{r}_{(\rm ee,hh)}$ and $\hat{r}_{\rm eh,he}$ are the normal reflection and AR matrices respectively, and $b_{0,x,y,z}$ and $a_{0,x,y,z}$ are the normal reflection and AR coefficients respectively. The AR in the channel $a_0$ always couples the electron and hole with opposite spin, which is independent of the direction of the spin quantization axis \cite{Suppl}, and therefore provides the spin-singlet channel. On the other hand, we can always find a spin basis, where the AR in one of the channels $a_{x,y,z}$ couples the electron and hole with the same spin \cite{Suppl} and therefore belongs to spin-triplet channels. As MFs have only spin-triplet superconducting correlations, we expect that the MF-induced resonant AR can occur only in spin-triplet channels. Below, we first use the scattering matrix theory \cite{TSC:Beenakker2011,TSC:Beenakker2014} to show the validity of our expectation at $E=0$ and then numerically explore the AR for the energy range inside the superconducting gap. For the D class TSC/NM interface, it was pointed out \cite{TSC:He2014} that the incoming electron and reflected hole in the MF-induced resonant AR process have the same spin. Therefore, this AR can occur only in spin-triplet channels. For a DIII class TSC/NM interface, the condition $\hat{T}\hat{R}^{\dagger}\hat{T}^{-1}=\hat{R}$, due to time-reversal symmetry, constrains $b_{x,y,z}=0$. Here, $\hat{T}=i\sigma_y K$. The unitarity of $\hat{R}$ further requires {\it either} 
$(|b_{0}|^2+|a_0|^2=1 \ \ {\rm and} \ \ a_x=a_y=a_z=0)$ {\it or} $(|a_x|^2+|a_y|^2+|a_z|^2=1 \ \ {\rm and} \ \ b_0=a_0=0)$ \cite{Suppl,TSC:Beenakker1997}.
The former corresponds to the trivial spin-singlet AR with $Q={\rm Pf} \left(i\sigma_y \hat{R}\right)=1$ \cite{TSC:Beenakker2011,TSC:Beenakker2014}. More importantly, the latter indicates a perfect spin-triplet AR and results in the nontrivial topological invariant $Q=-1$ \cite{TSC:Beenakker2011,TSC:Beenakker2014}. 

To further explore the AR inside the entire superconducting gap, we perform numerical calculations of ARs in two concrete 1D TSC tight-binding models. The Hamiltonian of the first TSC model, which violates time-reversal symmetry, has the form \cite{TSC:Sau2010,TSC:Lutchyn2010}
\begin{eqnarray}\label{Ham-SOC-1}
H_{\rm TS}&=&\left [-2t\cos(k)-\mu_{\rm s} \right ] \tau_z\otimes \sigma_0-M \tau_0\otimes \sigma_z\nonumber \\
&+&2t_{\rm so} \sin(k) \tau_z\otimes \sigma_y+\Delta \tau_x\otimes \sigma_0,
\end{eqnarray}
where $t$ is the spin independent hopping, $t_{\rm so}$ is the SOC hopping and $\mu_{\rm s}$ is the chemical potential of the SOC wire with $M$ the Zeeman coupling strength and $\Delta$ the proximity induced superconducting gap. The Hamiltonian of the normal metal is
\begin{eqnarray}\label{Ham-SOC-2}
H_{\rm NM}=(-2t\cos(k)-\mu_{\rm n})\tau_z\otimes \sigma_0,
\end{eqnarray}
where $\mu_{\rm n}$ is the chemical potential.
We numerically calculate the scattering matrix at $x=x_1$ (shown in Fig.~\ref{TSC-D}(a)) based on the Fisher-Lee relation \cite{Fisher1981a}, 
\begin{eqnarray}\label{F-L-1}
 \hat{R}_{ij}(E,\bm r)=-\delta_{ij}+i\hbar\sqrt{v_i v_j}G_{ij}^{\rm R}(E,\bm r),
 \end{eqnarray}
 where $G^{\rm R}$ is the retarded Green's function with $i,j=1,\dots,4$ and $v_{i(j)}$ is the velocity of the particle at energy $E$ in $i(j)$ channels.
In Fig.~\ref{TSC-D}(b), by adding a barrier potential $U=2t$ at the TSC/NM interface and taking $M=0.8t$, $\Delta=t_so=0.1t$, we plot the calculated conductance as a function of energy. The quantized zero-bias conductance with height $2e^2/h$ confirms the existence of a MF at the TSC/NM interface. Moreover, we plot the absolute value of the AR probabilities $|a_{0,x,y,z}|^2$ in Fig.~\ref{TSC-D}(c). It is shown that inside the gap, the AR only occurs in the spin-triplet channels $a_x$ and $a_y$, which is consistent with our prediction. In Figs.~\ref{TSC-D}(d) and \ref{TSC-D}(e), we also plot the conductance and the AR probabilities when the superconductor is in the trivial regime by choosing $M=\Delta/2=0.05t$. The conductance has no peak at $E=0$ and the AR only occurs in the spin-singlet channel. These results suggest that the perfect AR in the spin-triplet channels is solely induced by the MF. 

\begin{figure}[ht]
\centering
\begin{tabular}{l}
\includegraphics[width=0.9\columnwidth]{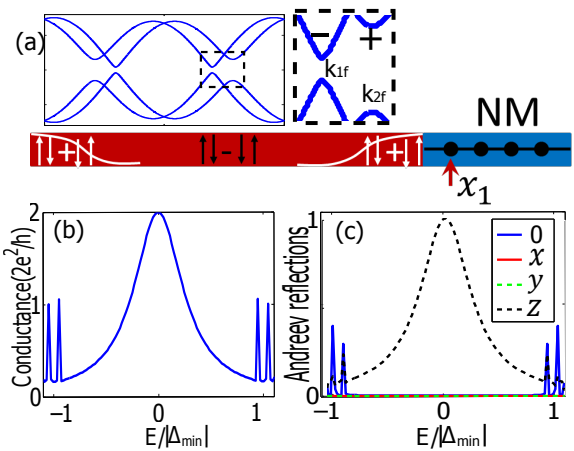}
\end{tabular}
\caption{(Color online) (a) Time-reversal invariant TSC/NM junction. The white arrows represent the spin-triplet superconducting correlations of MFs. The black arrows represent the spin-singlet bulk superconducting gap. The bulk dispersion of the TSC shows two superconducting gaps. The inset shows the sign of the minimal gap $\Delta_{\rm min}$ is negative. (b) The conductance of the TSC/NM junction as a function of energy. (c) AR probabilities for spin-singlet and spin-triple channels as a function energy.  The solid blue, solid red, dashed green and dashed black lines represent AR probabilities, $|a_0|^2$, $|a_x|^2$, $|a_y|^2$ and $|a_z|^2$ respectively.}
\label{TSC-DIII}
\end{figure}

Our prediction of the spin-triplet MF superconducting correlations is universal for both time-reversal violated and respected superconductors. Therefore, in the following, we choose a time-reversal symmetric TSC model whose Hamiltonian has the form \cite{TSC:Zhang2013}
\begin{eqnarray}\label{gap-iron}
H_{\rm TS}&=&\left [-2t\cos(k)-\mu_{\rm s}\right ] \tau_z\otimes \sigma_0 \nonumber \\
&&+2t_{\rm so} \sin(k) \tau_z\otimes \sigma_z+\Delta(k) \tau_x\otimes \sigma_0,
\end{eqnarray}
where $\Delta(k)=\left [\Delta_0-\Delta_1\cos(k) \right ]$, and is closed at $\cos(k_0)=\Delta_0/\Delta_1$. The bulk superconducting gap in Eq.~(\ref{gap-iron}) is a combined $s_{\pm}$ pairing potential and is obviously spin-singlet. The SOC in the semiconductor nanowire induces the spin splitting which results in two Fermi wave vectors $k_{\rm 1f}$ and $k_{\rm 2f}$. We assume $k_{\rm 1f}<k_{\rm 2f}$, as shown in the inset of Fig.~\ref{TSC-DIII}(a). It was pointed out \cite{TSC:Zhang2013} that for $k_{\rm 1f}<k_0<k_{\rm 2f}$,
the system is in the topological superconducting regime. In the following, we choose $\Delta_1=2\Delta_0=0.2t$, $\mu_{s}=-t$, and $t_{\rm so}=0.3t$. The gap sign is negative at $k=k_{\rm 1f}$ and positive at $k=k_{\rm 2f}$ as shown in Fig.~\ref{TSC-DIII}(a). We numerically calculate the reflection matrix at $x_1$ (Fig.~\ref{TSC-DIII}(a)). The conductance extracted from the reflection matrix is plotted as a function of energy in Fig.~\ref{TSC-DIII}(b). The quantized $4e^2/h$ zero bias-peak indicates that there is a pair of MFs at the end of the time-reversal invariant TSC. More importantly, we plot the AR probabilities $|a_{0,x,y,z}|^2$ at $x_1$ as a function of the energy in Fig.~\ref{TSC-DIII}(c). Inside the gap, we find that the AR occurs only in the spin-triplet channel $a_z$. The underlying physics is that the superconducting gaps at $\pm k_{\rm 1f}$ and $\pm k_{\rm 2f}$ have opposite signs. Therefore the AR coefficient associated with $\pm k_{1\rm f}$ also has the opposite sign to that associated with $\pm k_{\rm 2f}$ and thus the AR matrix is in the spin-triplet channel with the form $a_z\sigma_z$ instead of the spin-singlet channel with the form $a_0\sigma_0$. Therefore, although the bulk superconducting gap is spin-singlet, the AR at the TSC/NM interface is in the pure spin-triplet channel.
\begin{figure}[htbp]
\centering
\begin{tabular}{l}
\includegraphics[width=0.8\columnwidth]{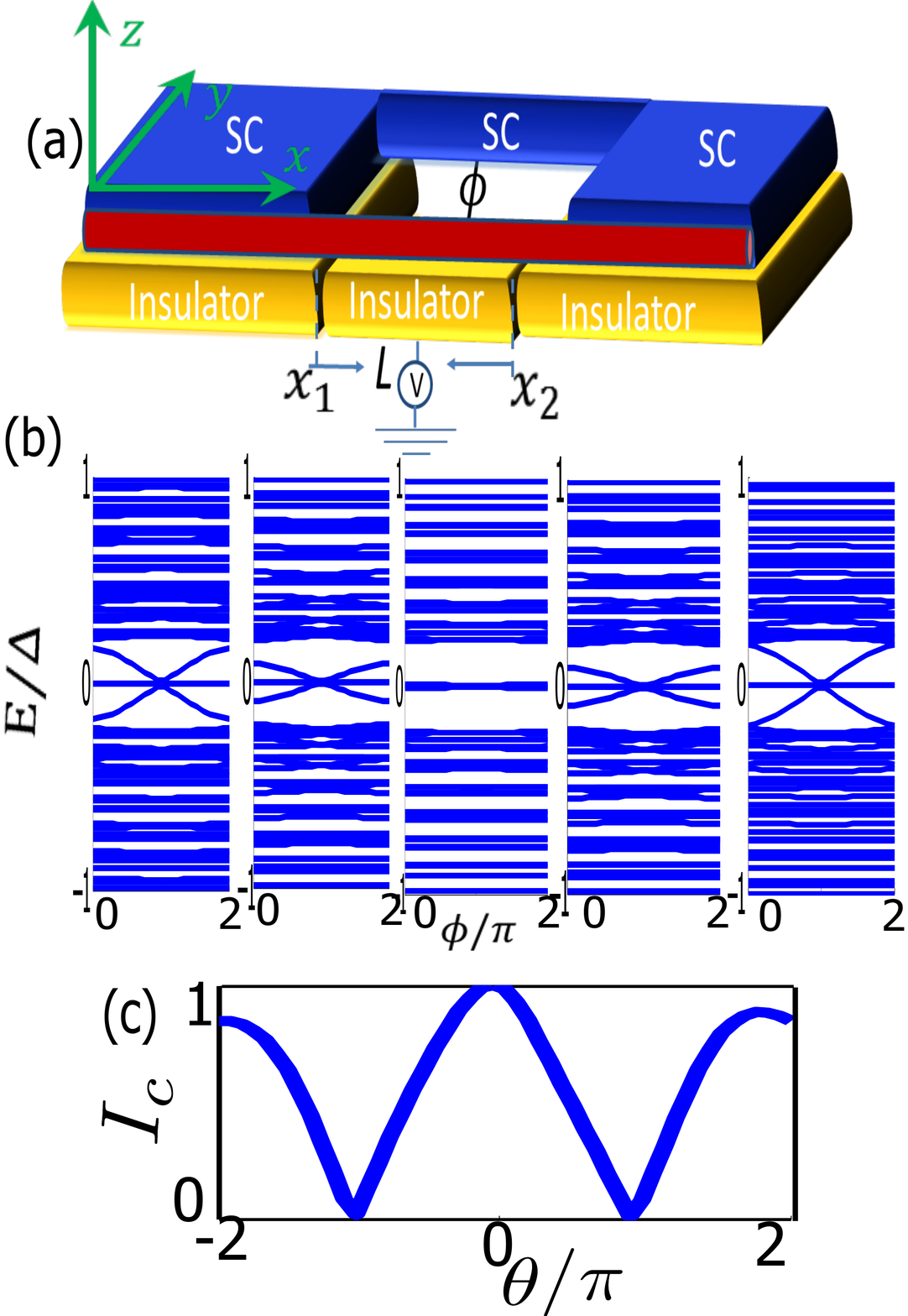}
\end{tabular}
\caption{(Color online) (a) TSC/SOC/TSC junction.   (b) Andreev levels for different SOC strengths which correspond to the precession angles $\theta=0, 3\pi/4, \pi, 5\pi /4, 2\pi$, respectively. (c) The oscillatory critical current as a function of $\theta$. The current is normalized by the critical current at $t'_{\rm so}=0$. }
\label{Andreev-Josephson}
\end{figure}

{\it SOC controlled critical Josephson current--} So far, we have shown that the MF-induced resonant AR occurs only in spin-triplet channels. It is known that the spin property of the AR-induced Cooper pairs is the same with that of the associated AR \cite{Liu2014}. Therefore, the MF-induced resonant AR always injects spin-triplet Cooper pairs. Recent studies \cite{TSC:Cheng2012,Bergeret2013a,Liu2014,Bergeret2014} show that SOC can rotate the \vec{d} vector of the AR-induced spin-triplet Cooper pairs but leave the spin-singlet pairs unaffected in the proximity region. Thus SOC provides a mechanism to distinguish between spin-triplet and spin-singlet Cooper pairs. To confirm our prediction, we therefore consider a TSC/SOC-semiconductor/TSC Josephson junction (Fig.~\ref{Andreev-Josephson}(a)). 
The Hamiltonian of the SOC-semiconductor system has the form
\begin{eqnarray*}
H_{\rm semi}=\left[-2t\cos(k)-\mu_{\rm s}\right]\tau_z\otimes \sigma_0+2t'_{\rm so} \sin(k) \tau_z\otimes \sigma_y,
\end{eqnarray*} where $t'_{\rm so}$ is the SOC hopping amplitude and the effective magnetic field of the SOC is along the $y$ direction. On the boundary $x_1$ and $x_2$ [Fig.~\ref{TSC-DIII}(a)] of time-reversal broken TSCs, based on our previous discussion, the Cooper pairs can be described as $|\uparrow\uparrow\rangle$, with spin $\uparrow$ along the $z$ direction. When a Cooper pair travels from $x_1$ to $x_2$ in the SOC-semiconductor region, because the effective SOC field is along the $y$ direction, the spins of the electron and hole start to precess in the $x$-$z$ plane with the same precession angle. As the spin splitting energy due to SOC is $\delta E=2t'_{\rm so}\sin(k_{\rm f})$, the precession angle of the particle traveling from $x_1$ to $x_2$ satisfies \cite{Suppl}
\begin{eqnarray}\label{precession}
\theta=\frac{\delta E L}{h v_{\rm f}}=N\frac{t'_{\rm so}}{t},
\end{eqnarray} where $v_{\rm f}\approx 2t\sin(k_{\rm f})$ in the limit $t'_{\rm so}\ll t$, and we assume there are $N$ sites in the normal SOC region whose length is $L$ (Fig.~\ref{Andreev-Josephson}(a)). The state of this Cooper pair at $x_2$ generally can be described as $|\nearrow\nearrow\rangle$ where $\nearrow$ represents a spin lying in the $x-z$ plane and its spin wave function is $(\cos(\theta/2), \sin(\theta/2))^{\rm T}$. Therefore, the projection of the condensate $|\nearrow\nearrow\rangle$ to $|\uparrow \uparrow \rangle$ takes the form
\begin{eqnarray}\label{projection}
\langle \uparrow \uparrow | \nearrow\nearrow\rangle=(1+\cos\theta)/2.
\end{eqnarray} Especially for $\theta=\pi$, the projection in Eq.~(\ref{projection}) is zero which means no Cooper pair can travel through the Josephson junction, and this indicates zero critical current. As a result, we expect the critical current will oscillate with a large amplitude and may even vanish as the SOC strength is suitably tuned. To confirm our theoretical argument, we plot the Andreev levels in Fig.~\ref{Andreev-Josephson}(b) in the cases of $\theta=0, 3\pi/4, \pi, 5\pi/4, 2\pi$ respectively. 
Because the Andreev levels cross $E=0$ at $\phi=\pi$, the Josephson current has the form $I_{\rm s}=I_{\rm c} \sin(\phi/2)$ where $I_{\rm c}$ is the critical current and $\phi$ is the phase difference between two TSCs. It is noted that at $\theta=\pi$, the energy-phase relation of the Andreev levels is almost flat. Therefore we conclude $I_c=0$ in this case which is consistent with our theoretical analysis. In Fig.~\ref{Andreev-Josephson}(c), $I_{\rm c}$ is plotted as a function of SOC controlled precession angle $\theta$, showing a $2\pi$ period as we expected based on Eq.~\ref{projection}. Generally, time-reversal broken TSCs can be considered as Cooper pair filters which only allow the Cooper pairs in the state $|\uparrow\uparrow\rangle$ to enter the TSCs from the normal region. However the situation for time-reversal invariant TSCs is very different. Because the superconducting condensates exist in both spin-up and spin-down channels, the spin-triplet Cooper pairs can always enter the time-reversal invariant TSCs which is independent of their spin direction \cite{Suppl}. 

Our prediction can also be applied to multi-band and even two-dimensional quantum wells with equal Rashba and Dresselhaus SOC strengths, which have been realized experimentally \cite{Weber:2007_a,Koralek:2009_a,Wunderlich2010,Yang2012,Walser2012,Dettwiler2014}. In this case, the spin precession angles for all transverse channels are the same \cite{Bernevig:2006_a,Stanescu:2007_a,LiuXin:2012_a} and the d-vector of the triple pairs will not decay even in the presence of spin-independent disorder \cite{Liu2014}. The oscillatory critical current has also been predicted and observed in the non-topological superconductor/ferromagnet/superconductor (SFS) Josephson junction where time-reversal is broken and MFs do not play any role in contrast to our system \cite{Bergeret2005a,Buzdin2005a}. The current-phase relation in the SFS junction undergoes a 0-$\pi$ transition when the critical current approaches zero. We emphasize that the current-phase relation of our time-reversal broken Josephson junction does not undergo a $0$-$\pi$ transition. This makes our prediction qualitatively different from that in the SFS junction. Therefore, our prediction of a SOC controlled critical current is a transport signature of MFs.
%
%

In a realistic experimental setup, the semiconductor wire may not necessarily be perfect 1D but may have several channels. However, our prediction requires the scattering at the TSC/NM interface to be 1D. The recent development of a surface split-gates design \cite{shabani:2014} can make a real 1D point contact between the topological superconducting semiconductor wire and the normal lead. Therefore, our experimental proposal can be realized by an existing experimental technique.

{\it Acknowledgment:} X.L. acknowledges useful discussions with Alejandro M. Lobos and Xiaopeng Li. This work is supported by Microsoft Q, JQI-NSF-PFC, and LPS-CMTC.


%

\begin{widetext}

\appendix{Supplementary information}

\section{Some knowledges }
\subsection{Particle-hole symmetry and time-reversal symmetry}

The BdG Hamiltonian, $\hat{H}$, of a superconductor has always particle-hole symmetry $\hat{C} \hat{H} \hat{C}=-\hat{H}$ where $\hat{C}$ is the particle-hole operator. In the basis $(c_{\uparrow},c_{\downarrow},c^{\dagger}_{\downarrow},-c^{\dagger}_{\uparrow})^{\rm T}$, the particle-hole operator is $\hat{C}=\tau_y \sigma_y K$ where $\tau$ and $\sigma$ are Pauli matrices in particle-hole and spin space respectively. The time-reversal operator has the form $\hat{T}=i\tau_0\sigma_y K$. If the system is time-reversal invariant, we have $\hat{T}^{-1} \hat{H} \hat{T}= \hat{H}$.

\subsection{Definition of spin-singlet and spin-triplet gaps}

Theoretically speaking, the Hamiltonian for superconducting pairing potential generally has the form
\begin{eqnarray}\label{gap-1}
H_{gap}=(\Delta_0 \Gamma_0+\tilde{\Delta}_0 \tilde{\Gamma}_0)+\sum_{i=1,2,3} (\Delta_j\Gamma_j+\tilde{\Delta}_j \tilde{\Gamma}_j),
\end{eqnarray}
where $\Delta_0(\tilde{\Delta}_0)$ is the real (imaginary) part of the s-wave gap, $\Delta_{j} (\tilde{\Delta}_j)$ is the imaginary (real) part of the p-wave gap, $\Gamma_{0,1,2,3}=(\tau_x \otimes \sigma_0,\tau_y\otimes \sigma_x, \tau_y\otimes \sigma_y ,\tau_y\otimes \sigma_z)$ and $\tilde{\Gamma}_{0,1,2,3}=(\tau_y \otimes \sigma_0,\tau_x\otimes \sigma_x, \tau_x\otimes \sigma_y ,\tau_x\otimes \sigma_z)$  in the basis $(c_{\uparrow},c_{\downarrow}, c^{\dagger}_{\downarrow},-c^{\dagger}_{\uparrow})^{\rm T}$. We find that the matrices $\Gamma_{0,1,2,3}$ and $\tilde{\Gamma}_{0,1,2,3}$ satisfy the Dirac algebra 
\begin{eqnarray}\label{Dirac-algebra-1}
\left\{ \Gamma_{\mu}, \Gamma_{\nu} \right \}=2I_{4\times 4}, \ \ \left\{ \tilde{\Gamma}_{\mu}, \tilde{\Gamma}_{\nu} \right \}=2I_{4\times 4}, \ \ \tilde{\Gamma}_{\mu=0,1,2,3}=\Gamma_{5}\Gamma_{\mu=0,1,2,3},
\end{eqnarray} where $\Gamma_5=i\Gamma_1\Gamma_2\Gamma_3\Gamma_4=-i\tau_z\sigma_0$. Therefore, the pairing potential can be constructed from these eight gamma matrices. In the following, we will prove that $\Gamma_{i=1,2,3}$ expand a three dimensional vector space and construct a three dimensional representation of a SU(2) group whose generators correspond to the rotation in spin space. 

First, we define 
\begin{eqnarray}\label{SU(2)-1}
\frac{i}{4}[\Gamma_{i},\Gamma_{j}]=\frac{1}{2}\epsilon_{ijk}\tau_0 \otimes \sigma_k=\frac{1}{2}\epsilon_{ijk}\left(\begin{array}{cc} \sigma_k & 0 \\ 0 &  \sigma_k \end{array}\right)=\epsilon_{ijk}S_k,
\end{eqnarray}
where $S_{i=1,2,3}$ are three generators of a SU(2) algebra.
Next, we will show that the matrix $e^{-i\bm S \cdot \bm n}$ corresponds to rotating the spin from z axis to the $\bm n$ direction where $\bm n$ is a three dimensional unit vector. Generally a wave function in the basis $(c_{\uparrow},c_{\downarrow},c^{\dagger}_{\downarrow},-c^{\dagger}_{\uparrow})^{\rm T}$ has the form
\begin{eqnarray}\label{wf-10}
\Psi=\left(\begin{array}{c} \psi_{\rm e} \\ i\sigma_y \psi_{\rm h} \end{array}\right).
\end{eqnarray}
where $\psi_{\rm e (h)}$ is the 2-component spinor wave function for the electron (hole). When we rotate the electron spin as
\begin{eqnarray}\label{wf-11}
 \psi'_{\rm e}=e^{-i\bm \sigma \cdot \bm n/2}\psi_{\rm e},
\end{eqnarray}
the spinor wave function for the hole will be transformed as
\begin{eqnarray}\label{wf-12}
\psi'_{\rm h}=e^{i\bm \sigma^* \cdot \bm n/2}\psi_{\rm h}.
\end{eqnarray}
Therefore the wave function is transformed as
\begin{eqnarray}\label{wf-13}
\Psi'=\left(\begin{array}{c} \psi'_{\rm e} \\ i\sigma_y \psi'_{\rm h} \end{array} \right)=\left(\begin{array}{c} e^{-i\bm \sigma \cdot \bm n}\psi_{\rm e} \\ i\sigma_y e^{i\bm \sigma^* \cdot \bm n /2}\psi_{\rm h} \end{array} \right)=\left(\begin{array}{c} e^{-i\bm \sigma \cdot \bm n /2}\psi_{\rm e} \\  e^{-i\bm \sigma \cdot \bm n} i\sigma_y\psi_{\rm h} \end{array} \right)=e^{-i\bm S \cdot \bm n} \Psi.
\end{eqnarray}
Therefore the matrix $e^{-i\bm S \cdot \bm n}$ corresponds to the rotation in the spin space. Because the SU(2) generators $S_{i=1,2,3}$ are constructed from the Eq.~\ref{SU(2)-1}, the three gamma matrices $\Gamma_{1,2,3}$ construct a three dimensional representation of a SU(2). Therefore $\Gamma_{1,2,3}$ are the spinnors of the SU(2) with generators $S_k=\epsilon_{ijk}\tau_0\sigma_k/2$. This is consistent to the fact the $\Gamma_{1,2,3}$ are the basis of spin-triplet pairing. It is easy to see that $(\Gamma_1,\Gamma_2)=(i\tau_y\sigma_x,i\tau_y\sigma_y)$ represent the superconducting pairing in the same spin channel. Therefore, as long as the superconducting pairing can be written as the superposition of the basis $\Gamma_{1,2,3}$, we can always find a rotation operator which can rotate a general spinor to the axis in which the electron and hole have the spin. Also it is easy to check that $\left[ \Gamma_0,S_{i=1,2,3} \right]=0$. Therefore $\Gamma_0$ is a one-dimensional representation of the SU(2) spin rotation group which is consistent to the fact that $\Gamma_0$ is the basis of the spin-singlet pairing. The similar results can be applied to $\tilde{\Gamma}_{0,1,2,3}$.
\section{Spectral function and pairing density}

The equal space spectral function at energy $E$ is defined as 
\begin{eqnarray*}
A(E,x)=\frac{i}{2\pi}(G^{R}(E,x)-G^{A}(E,x)),
\end{eqnarray*} 
which is a Hermitian matrix.
 Based on the Lehmann representation, the retarded Green's function is related to the spectral function  as 
\begin{eqnarray}\label{GF-1}
G^R(E,x)=(E-H+i\delta)^{-1}&=&\int _{-\infty}^{\infty}  \frac{A(\epsilon,x)}{E-\epsilon+i\delta} d\epsilon ={\rm P} \int_{-\infty}^{\infty} \frac{A(\epsilon,x)}{E-\epsilon}d\epsilon-i\pi A(E,x) \nonumber,
\end{eqnarray}
 where $x$ and $x'$ are the spatial coordinates. 
Because the spectral function $A(\epsilon,x)$ is hermitian, the principle integral ${\rm P} \int_{-\infty}^{\infty} \frac{A(\epsilon,x)}{E-\epsilon}$ is hermitian, and $-i\pi A(E,x)$ is anti-hermitian. The spectral function can be written as $A(\epsilon)=\sum_{\epsilon_n=\epsilon}|\Psi_n(x)\rangle\langle \Psi_n(x)|$. As we are interested in the superconducting pairing, in the following, we will focus on the off diagonal part of the spectral function which has the form $A^{\rm off}(\epsilon)=d_{\mu}(\epsilon,x)\Gamma_{\mu}+\tilde{d}_{\mu}(\epsilon,x)\tilde{\Gamma}_{\mu}$ where $d_{\mu}$ and $\tilde{d}_{\mu}$ are the superconducting pairing amplitudes. Because $\Gamma_{\mu}$, $\tilde{\Gamma}_{\mu}$ and the spectral function $A(\epsilon,x)$ are hermitian, the coefficients $d_{\mu}$ and $\tilde{d}_{\mu}$ are real. We now focus on the retarded Green's function at $E=0$ because it is related to the MFs in the topological superconductors. As the Hamiltonian $H$ has particle-hole symmetry, the retarded Green's function at $E=0$ also respects particle-hole symmetry which is shown below:
\begin{eqnarray}\label{GF-2}
\hat{C} G^{\rm R}(0,x) \hat{C}^{-1}&=&\hat{C} (-H+i\delta)^{-1} \hat{C}^{-1}=-(-H+i\delta)^{-1}\equiv- {\rm P}\int_{-\infty}^{\infty} \frac{A(\epsilon,x)}{-\epsilon} d\epsilon+i\pi A(E,x)\end{eqnarray}
Particularly we are interested in the block off diagonal part of the retarded Green's function
\begin{eqnarray}\label{GF-anoma}
F^{\rm R}=\left(\begin{array}{cc} 0& f^{\rm R} \\ \overline{f}^{\rm R} & 0\end{array}\right).
\end{eqnarray}
Based on Eq.~\ref{GF-2}, the off diagonal retarded Green's function satisfies
\begin{eqnarray}\label{GF-4}
\hat{C} F^{\rm R}(0,x) \hat{C}^{-1}&=&{\rm P}\int_{-\infty}^{\infty} \frac{A^{\rm off}(\epsilon,x)}{\epsilon} d\epsilon+i\pi A^{\rm off}(0,x)\nonumber \\ &=& {\rm P}\int_{-\infty}^{\infty} \frac{d_{\mu}(\epsilon,x)\Gamma_{\mu}+\tilde{d}_{\mu}(\epsilon,x)\tilde{\Gamma}_{\mu}}{\epsilon} d\epsilon+i\pi \left(d_{\mu}(0,x)\Gamma_{\mu}+\tilde{d}_{\mu}(0,x)\tilde{\Gamma}_{\mu}\right).
\end{eqnarray}
On the other hand, particle-hole symmetry requires the anomalous Green's function to satisfy 
%
\begin{eqnarray}\label{GF-5}
 &&\hat{C} F^{\rm R}(0,x) \hat{C}^{-1}=\hat{C} \left({\rm P} \int_{-\infty}^{\infty} \frac{A^{\rm off}(\epsilon,x)}{-\epsilon}-i\pi A^{\rm off}(0,x)\right) \hat{C}^{-1} \nonumber \\
&&= -{\rm P} \int_{-\infty}^{\infty} \frac{d_{0}(\epsilon,x)\Gamma_{0}+\tilde{d}_{0}(\epsilon,x)\tilde{\Gamma}_{0}-d_{i}(\epsilon,x)\Gamma_{i}-\tilde{d}_{i}(\epsilon,x)\tilde{\Gamma}_{i}}{\epsilon}+i\pi \left(-d_{0}(0,x)\Gamma_{0}-\tilde{d}_{0}(0,x)\tilde{\Gamma}_{0}+d_{i}(0,x)\Gamma_{i}+\tilde{d}_{i}(0,x)\tilde{\Gamma}_{i}\right),\nonumber \\
\end{eqnarray} 
where we use the fact that
\begin{eqnarray*}
\hat{C} \Gamma_{\mu=0,1,2,3} \hat{C}^{-1}=(-\tau_x \sigma_0, \tau_y\sigma_x,\tau_y\sigma_y,\tau_y\sigma_z),  \ \ \ \hat{C} \tilde{\Gamma}_{\mu=0,1,2,3} \hat{C}^{-1}=(-\tau_y\sigma_0,\tau_x\sigma_x,\tau_x\sigma_y,\tau_x\sigma_z).
\end{eqnarray*}

Comparing Eq.~\ref{GF-4} with Eq.~\ref{GF-5}, we conclude that spin-singlet pairing is completely forbidden in the anti-hermitian part of the anomalous Green's function $F^{\rm R}_{anti-hermitian}(0,x)=-i\pi A^{\rm off}(0,x)$ which indicates that the MFs cannot have spin-singlet pairing.



\section{Scattering matrix theory for DIII topological superconductor}

When a normal lead is attached to a superconductor, the Andreev reflection at the SC/NM interface will induce the superconducting condensate in the normal region. In the following and for a one-dimensional (1D) SC/NM junction, we use scattering matrix theory to show that the induced superconducting condensate by the MF resonant Andreev reflection is always spin-triplet. 
We first focus on a DIII class SC/NM junction. Because of time-reversal symmetry, the scattering matrix is self dual \cite{TSC:Beenakker1997} which requires that
\begin{eqnarray*}
S=\left(\begin{array}{cc} r_{ee} \sigma_0 & a_0\sigma_0+a_j\sigma_j \\
a\sigma_0-a_j\sigma_j & r_{\rm hh}\sigma_0 \end{array} \right).
\end{eqnarray*}
It is straightforward to show that the above scattering matrix satisfies the condition $\hat{T}S^{\dagger}(E)\hat{T}^{-1}=S(E)$. It is also noted that $a_{j=x,y,z}$ are the spin-triplet scattering amplitudes and $a_0$ is the spin-singlet scattering amplitude. We are interested in the Andreev reflection at $E=0$. In this case, according to the particle-hole symmetry of the system, we have
\begin{eqnarray*}
\tau_y\sigma_y S^* \tau_y\sigma_y &=& \left(\begin{array}{cc} 0 & -i\sigma_y \\
i\sigma_y & 0 \end{array} \right) \left(\begin{array}{cc} r_{ee} \sigma_0 & a_0\sigma_0+a_j\sigma_j \\
a\sigma_0-a_j\sigma_j & r_{\rm hh}\sigma_0 \end{array} \right) \left(\begin{array}{cc} 0 & -i\sigma_y \\
i\sigma_y & 0 \end{array} \right) \nonumber \\
&=&\left(\begin{array}{cc} r^*_{\rm hh} \sigma_0 & -(a_0^*\sigma_0+a_j^*\sigma_j) \\
-(a^*\sigma_0-a_j^*\sigma_j) & r_{ee}^*\sigma_0 \end{array} \right)\nonumber \\
&=&S=\left(\begin{array}{cc} r_{\rm ee} \sigma_0 & a_0\sigma_0+a_j\sigma_j  \\
a_0\sigma_0-a_j\sigma_j  & r_{\rm hh}\sigma_0 \end{array} \right).
\end{eqnarray*}
Therefore, we have $r_{ee}^*=r_{\rm hh}=b_0$ is real, $a_0$ and $a_j$ are pure imaginary. At last, the matrix $S$ should be unitary. This requires
\begin{eqnarray*}
S^{\dagger}S=\left(\begin{array}{cc}  \sigma_0 & 0 \\
0 & \sigma_0 \end{array} \right)
\end{eqnarray*}
Because
\begin{eqnarray*}
(S^{\dagger}S)_{\rm ee}=(|b_0|^2+|a_0|^2+|a_x|^2+|a_y|^2+|a_z|^2)\sigma_0+2a_0 a_j\sigma_j=\sigma_0,
\end{eqnarray*}
we have either $a_{j=x,y,z}=0$, $|b_0|^2+|a_0|^2=1$ or $a_0=0$, $|b_0|^2+|a_x|^2+|a_y|^2+|a_z|^2=1$. These indicate that the spin-singlet Andreev reflection (the former case) and spin-triplet Andreev reflection (the latter case) cannot coexist. For the spin-triplet Andreev reflection, we further have
\begin{eqnarray*}
(S^{\dagger}S)_{\rm eh}=2b_0a_j\sigma_j=0,
\end{eqnarray*}
which requires either $r^*=0$ or $a_{j=x,y,z}=0$ and indicates that the normal reflection and spin-triplet Andreev reflection also cannot coexist. Therefore, we conclude for a DIII class SC/NM junction and at $E=0$, the spin-triplet Andreev reflection must be a perfect Andreev reflection with vanished spin-singlet component. It is also straightforward to show that 
\begin{eqnarray*}
\rm{Pf} \left( i\sigma_y\tau_0 \left(\begin{array}{cc} 0 &  a_j\sigma_j \\
- a_j\sigma_j & 0 \end{array} \right) \right)&=&\rm{Pf} \left(\begin{array}{cc} 0 & i\sigma_y a_j\sigma_j \\
-i\sigma_y a_j\sigma_j & 0 \end{array} \right) \\
&=&-(|a_x|^2+|a_y|^2+|a_z|^2)=-1,
\end{eqnarray*}
which is consistent to the condition of the nontrivial topological quantum number \cite{TSC:Beenakker2014} in terms of the scattering matrix for a DIII class superconductor. 

\section{Spin precession angle and its effect for time-reversal invariant topological junction}
The Hamiltonian of the SOC-semiconductor is
\begin{eqnarray}\label{Ham-SOC-semi-1}
H_{\rm semi}=(-2t\cos(ka)-\mu_{\rm s})\tau_z\otimes \sigma_0+2t'_{\rm so} \sin(ka) \tau_z\otimes \sigma_y,
\end{eqnarray} where $k$ is the wave vector, $a$ is the lattice constant, $t'_{\rm so}$ is the SOC hopping constant in the SOC-semiconductor and the effective magnetic field of the SOC is along $y$ direction. SOC will shift the two spin-subbands to opposite direction in the momentum space. In the 1D tight-binding model with $k_x\sigma_y$-like SOC Hamiltonian at low energy limit, the dispersion relation is $E=2t(1-\cos(ka))\pm 2t_{\rm so}\sin(ka)=2t-2\sqrt{t^2+t_{\rm so}^2}\cos(ka\pm \delta \phi)$ where $\cos(\delta \phi)=t/\sqrt{t^2+t_{\rm so}^2}$. If there are N sites in the normal region, the electron and hole will take time $T=N/ v_f \approx N\hbar/2at\sin(k_f a)$ to go through the normal region. Because the effective magnetic field due to SOC is $B_{\rm soc}=2t'_{\rm so} \sin(k_f a)$, the spin precession angle of an electron or hole in the normal region is $\gamma=B_{\rm soc} T=Nt'_{\rm so}/t$. In the continuous limit, we have $t\rightarrow\hbar^2/2ma^2$, $t'_{\rm so}\rightarrow \alpha/a$. Therefore, the spin precession angle in the continuous limit is $\gamma=Nt'_{\rm so}/t=2m\alpha L$ where $L=Na$ is the length of the normal region.

For a time-reversal invariant TSC, there is a pair of Majorana zero modes with opposite spins at the boundary. The induced spin-triplet condensates are described by  $(|\uparrow\uparrow\rangle+|\downarrow\downarrow\rangle)/\sqrt{2}$ which is even under time-reversal operation. Here we assume the Majorana spins are parallel or anti-parallel to z axis. If the spin-orbit coupling (SOC) field direction is along y axis as we consider in the manuscript, the spins are rotated so that the spin-triplet states are changed to $(|\nearrow\nearrow\rangle+|\swarrow\swarrow\rangle)/\sqrt{2}$ where $\nearrow$ and $\swarrow$ represent the spins lying in the x-z plane with wave functions $(\cos(\theta/2),\sin(\theta)/2)^{\text T}$ and $(\sin(\theta/2),-\cos(\theta)/2)^{\text T}$ respectively. Therefore the projection of the condensates is
\begin{eqnarray*}
(\langle \uparrow\uparrow|+\langle \downarrow\downarrow|)(|\nearrow\nearrow\rangle+|\swarrow\swarrow\rangle)/2=1,
\end{eqnarray*}
which is independent of the SOC. Therefore in this case the TSC/SOC/TSC junction performs like a normal Josephson junction. However if the SOC field direction is along x axis so that the Majorana spin is rotated in y-z direction, the spins $\nearrow$ and $\swarrow$ correspond to $(\cos(\theta/2),i\sin(\theta)/2)^{\text T}$ and $(i\sin(\theta/2),\cos(\theta)/2)^{\text T}$ respectively. The projection is  
\begin{eqnarray*}
(\langle \uparrow\uparrow|+\langle \downarrow\downarrow|)(|\nearrow\nearrow\rangle+|\swarrow\swarrow\rangle)/2=\cos\theta,
\end{eqnarray*}
which will result in a 0- and $\pi$- Josephson junction transition.

%

\end{widetext}

\end{document}